\documentclass[%
 aip,
 amsmath,amssymb,
 reprint,%
]{revtex4-1}

\usepackage{graphicx}
\usepackage{dcolumn}
\usepackage{bm}

\usepackage[utf8]{inputenc}
\usepackage[T1]{fontenc}
\usepackage{mathptmx}

\usepackage{float} 

\usepackage{physics}
\usepackage{braket}
\usepackage{mhchem}
\usepackage{color}

\usepackage{tabularx}
\usepackage{siunitx}
\usepackage{xcolor}

\usepackage{textgreek}

\begin{document}

\preprint{AIP/123-QED}

\title{A Promising Intersection of Excited-State-Specific Methods
       from Quantum Chemistry and Quantum Monte Carlo}

\author{Leon Otis}
\affiliation{Department of Physics, University of California Berkeley, CA 94720,USA }
\author{Eric Neuscamman}
\email{eneuscamman@berkeley.edu}
\affiliation{Department of Chemistry, University of California Berkeley, CA 94720,USA}
\affiliation{Chemical Sciences Division, Lawrence Berkeley National Laboratory, Berkeley, CA, 94720, USA}

\date{\today}

\begin{abstract}

We present a discussion of recent progress in excited-state-specific
quantum chemistry and quantum Monte Carlo alongside a demonstration
of how a combination of methods from these two fields can offer
reliably accurate excited state predictions across singly excited,
doubly excited, and charge transfer states.
Both of these fields have seen important advances supporting excited
state simulation in recent years, including the introduction of
more effective excited-state-specific optimization methods,
improved handling of complicated wave function forms,
and ways of explicitly balancing the quality of wave functions
for ground and excited states.
To emphasize the promise that exists at this intersection,
we provide demonstrations using a combination of
excited-state-specific complete active space self-consistent field theory, 
selected configuration interaction, and state-specific variance minimization.
These demonstrations show that combining excited-state-specific
quantum chemistry and variational Monte Carlo can be more reliably accurate
than either equation of motion coupled cluster theory or multi-reference
perturbation theory, and that it can offer new clarity in
cases where existing high-level methods do not agree.

\end{abstract}

\maketitle

\section{Introduction}

\subsection{Challenges Faced by Current Methods}

The wide variety of different excited states in molecules poses an ongoing challenge for traditional 
electronic structure methods.
Many of the most widely applied and scalable methods may provide satisfyingly accurate 
predictions for one class of excited states, only to fail on others.
When surveying the landscape of electronic structure techniques, this problem of inconsistent
performance can be traced to the difficulty of simultaneously satisfying the multiple requirements 
for accurate treatment of all classes of excited states.
First, a method should have the means of correctly describing multiconfigurational character, which 
is more common among excited states compared to ground states.
At the same time, for quantitatively accurate energies, the method must possess an 
ability to capture weak correlation effects.
This effective treatment of strong and weak correlation should be maintained in a balanced fashion 
for both ground and excited state in order to obtain correct energy differences, and accounting 
for orbital relaxations may be necessary for cases where orbital shapes differ significantly between 
the two states.
Any potential method with general-purpose applicability to excited states must also satisfy the above 
considerations in larger molecules and basis sets.
After discussing the individual strengths and weaknesses of some widely
used quantum chemistry approaches, we will review recent advances in
excited-state-specific methods in variational Monte Carlo and
quantum chemistry and explore the promise offered by their combination.

Time-dependent density functional theory (TDDFT) is by far the most commonly used approach for 
excited states and has played a leading role in the modeling of absorption spectra.\cite{Casida2012}
However, TDDFT has historically struggled, at least for its most common functional choices within the 
adiabatic approximation, to describe some classes of excited states accurately,
with double excitations and charge transfer (CT) as two notorious
examples.\cite{Maitra2004,Levine2006,Dreuw2003,Dreuw2004}
While excited state DFT methods continue to benefit from extensive 
research\cite{Elliott2011,Maitra2017,Hait2020a,Hait2021} analyzing and improving 
approximate functionals, these categories of excited states remain challenging.
One alternative approach is to employ Green's function techniques such as GW-BSE, which 
can obtain improved excitation energies over TDDFT,\cite{Bruneval2015,Jacquemin2015,Rangel2017} but these methods in practice rely on an
initial 
set of DFT eigenvalues and eigenstates and are sensitive to this choice of starting point.
These challenges, and the ever-growing 
array \cite{Mardirossian2017,Laurent2013} of possible functional choices,
motivate the development and use of methods with greater
predictive power for difficult categories of excited states.

Among wave function approaches to excited states, coupled cluster theory has
much to recommend it, given its prowess for ground state properties
and its effective treatment of weak correlation.
\cite{Bak2000,Coriani2005,Bartlett2007}
Excited state extensions of coupled cluster, particularly the equation-of-motion
formulation \cite{Stanton1993} (EOM-CC), have become widely used at different levels of cluster truncation. 
For single valence excitations, limiting the excitation level to singles and doubles with EOM-CCSD 
obtains balanced results with a systematic error of only a few tenths of an eV.\cite{Kannar2014a}
Approaches which include higher orders can achieve even better accuracy. Notably, the approximate inclusion of triple excitations, as in iterative models 
such as CC3, is sufficient to achieve comparable accuracy to EOM-CCSDT (a few hundredths of an eV) for this category of excitations.\cite{Loos2018}

However, for double excitations, EOM-CCSD produces substantially
larger errors than for single excitations \cite{Hirata2000} and,
while higher orders of coupled cluster can obtain accurate results 
in some cases, \cite{Watson2012} their errors can still be as
large as half an eV or worse for pure double excitations. \cite{Loos2019}
Besides these inaccuracies, the cost-scalings of high-level coupled cluster methods
($N^8$ for EOM-CCSDT and $N^{10}$ for EOM-CCSDTQ) hinders their application
to large molecules, even for types of excited states for which they are well suited.

The difficulties faced by coupled cluster theory in strongly correlated states
like many double excitations helps motivate the use of multi-reference
approaches such as complete active space second order perturbation theory (CASPT2).
\cite{SerranoAndres1993,SerranoAndres1994,Nakayama1998,Ostojic2001,Dallos2004,Schreiber2008,Silva-Junior2010,Duman2012,Wen2018,Rauer2016,BenAmor2017,Loos2019,Loos2020}
While CASPT2 faces its own limitations due to intruder states,
sensitivity to active space and state-averaging choices, and the
limiting of its weak correlation treatment to the PT2 level, it is still
capable of obtaining accurate excitation energies, most notably in
double-excitation-dominated states where coupled cluster struggles. \cite{Loos2019}
In CT states, however, the situation is reversed,
with CASPT2's accuracy suffering \cite{Domingo2012,Meyer2014,Tran2019}
as state-averaging \cite{Werner1981} prevents the orbitals it uses
from fully relaxing in the presence of the charge transfer.
Indeed, recent efforts to build accurate benchmark sets for CT states, which are 
currently less common than for other classes of excited states,
have relied upon coupled cluster theory. \cite{Dutta2018,Kozma2020,Loos2021}
Whether it is coupled cluster's difficulties with double excitations or
CASPT2's difficulties with charge transfer or TDDFT's difficulties with both,
the limitations of the most widely used excited state methods motivate the
development of approaches that can offer accurate treatments regardless of
which type of excited state is being considered.

While a number of methods, \cite{Eriksen2020,Eriksen2021}
including selected configuration interaction (sCI),
\cite{Schriber2017,Holmes2017,Chien2018,Loos2018,Loos2019}
density matrix renormalization group (DMRG), \cite{Legeza2003,Sharma2014a}
and full configuration interaction quantum Monte Carlo (FCIQMC),
\cite{Booth2012,Humeniuk2014,Blunt2015a,Blunt2015b,Blunt2017b}
can achieve this reliability with systematic exploitation of wave
function sparsity and/or compressibility, their exponential scaling 
limits their prospects for extensions beyond small systems.
A recent study \cite{Eriksen2020} of benzene is a useful demonstration of
the state-of-the-art capabilities of this collection of high accuracy techniques.
The determination of benzene's ground state energy in a double-$\zeta$ basis
set is both an important  example of the molecular sizes that these high-level
approaches can now treat and a reminder of the challenges they face in larger
systems and the larger basis sets needed for excited state accuracy.
So far, the role of sCI, DMRG, and FCIQMC in larger molecules beyond benzene
has been limited to mostly serving as (impressively large) active space solvers.
\cite{Kurashige2013,Liu2013,Sharma2014b,Levine2020,Dobrautz2021}
Thus, whether one looks to coupled cluster, multi-reference perturbation theory, or the
modern array of sparsity and compressibility approaches, it remains challenging to deliver
consistently accurate excitation energies in larger molecules, larger basis sets,
and across the wide variety of excitation types.

In this Focus Article, we will discuss advances in excited-state-specific
methods in quantum chemistry (QC) and quantum Monte Carlo (QMC)
and the opportunities offered by their intersection.
Variational Monte Carlo \cite{toulouse2016,rubenstein2017,becca2017} (VMC)
in particular has seen a great deal of progress in modeling
excited states in recent years. \cite{feldt2020}
Thanks to improved methods for handling large Slater determinant
expansions and performing wave function optimization, excited
state VMC is now in a much stronger position to take full advantage
of recent advances in sCI and excited-state-specific quantum chemistry.
As has long been true in ground state simulations, it is increasingly
possible for excited state theories from quantum chemistry to partner
with quantum Monte Carlo to deliver systematically improvable
descriptions of excited states.
When these two methodologies work together in a fully
excited-state-specific manner,
it is possible to handle strong correlation, weak correlation,
and post-excitation orbital relaxations in a way that leads to
reliable accuracy for singly excited,
doubly excited, and charge transfer states in system sizes beyond
what is currently possible for DMRG, sCI, and FCIQMC.
To see how these advantages come about, let us
begin by overviewing the progress that has made them
possible before discussing the QC/QMC excited state
intersection more directly.

\subsection{Recent Developments in Variational Monte Carlo}

\subsubsection{Optimization}

Whether aiming to describe a ground or excited state, the optimization
of sophisticated wave function approximations is a key challenge in VMC
that has seen substantial progress in recent years.
In the early years of VMC optimization, it was often only a handful of
Jastrow factor parameters that were optimized, with the remaining parameters
(such as those that control orbital shapes) typically left at whatever
values had been determined by DFT or quantum chemistry.
\cite{Fahy1988,Fahy1990,Williamson1996,Foulkes2001}
Progress on methods that incorporate second-derivative
information --- such as modified Newton, \cite{umrigar2005}
the linear method
\cite{Nightingale2001,Umrigar2007,Toulouse2007,Toulouse2008}
(LM), and stochastic reconfiguration
\cite{Sorella2001,Casula2004,Sorella2005,Sorella2007}
(SR) --- made full wave function optimization,
including Jastrow, orbital, configuration interaction, and
pairing function parameters, more routine.
\cite{Casula2004,Sorella2005,Sorella2007,Umrigar2007,Toulouse2007,Toulouse2008,Cordova2007,Lawson2008,Zaccheddu2008,Filippi2009,Valsson2010,Send2011,Filippi2012,Fracchia2012}
However, most optimizations were limited to hundreds of parameters,
which stands in stark contrast to the thousands and millions of parameters
involved in quantum chemistry methods like coupled cluster theory
or recently-introduced neural-network wave functions for VMC.
\cite{carleo2017,pfau2020,hermann2020,choo2020,schatzle2021}
LM and SR optimizations of large parameter sets (which we will define
as those beyond a few thousand parameters) have to contend with two
significant challenges:
the memory required to store these methods' matrices and
the need to take large enough samples so that the statistical
uncertainty in these matrices' elements does not spoil the optimization.

Consider the memory issue first.  
In the original formulation of both the LM and SR, one constructs the
overlap matrix in the basis of the wave function's first derivatives with
respect to optimizable parameters
(the LM also constructs the Hamiltonian in this basis).
Storage for these matrices grows quadratically with dimension, 
leading for example to quartic memory growth with system size for
parameter sets (like the elements of the orbital coefficient matrix
and the parameters of some Jastrow factors) that themselves grow
quadratically in number with system size.  
The last decade has seen three promising developments that each allow
for substantial reductions in memory footprint.
First, it is possible to employ Krylov-subspace-based algorithms
to solve the linear equation in SR or the eigenvalue equation
in the LM without ever constructing the matrices in question.
\cite{Neuscamman2012,sabzevari2020}
Second, accelerated descent (AD) methods adapted from the machine learning
community allow for optimization without involving these matrices
at all, \cite{Schwarz2017,Sabzevari2018,Luo2018,Mahajan2019}
and, although AD methods are typically not as effective at
driving an optimization close to the exact minimum, \cite{Otis2019}
they can be applied to far larger parameter sets with less concern for
the effects of statistical uncertainty.
Third, the memory footprint can be reduced by
methods that approximate either SR or the LM in ways that allow them
to avoid having to store or even evaluate all the matrix elements.
Examples of this approach include multi-stage blocking \cite{Zhao2017}
for the LM and variants of the Kronecker-factored approximate curvature
(KFAC) method that are closely related to SR. \cite{pfau2020}
We have found that hybrids that combine AD methods with approximated
matrix methods are especially effective, \cite{Otis2019,Otis2020}
and indeed this hybrid approach is used in the VMC optimization stage
for the results presented later in this article.

Turning to the somewhat less straightforward issue of statistical
uncertainty, it is important to recognize that the key working equations
of the LM and SR involve either the diagonalization or inversion of
a matrix and that, when the matrix dimension becomes large,
the results of these operations are highly nonlinear functions of
the matrix elements.
Indeed, a matrix's eigenvalues come from its characteristic
polynomial, whose degree of nonlinearity is the matrix dimension itself.
Thus, the basic expectation that the LM and SR will, as Monte Carlo methods,
produce statistically uncertain optimization steps is compounded by the
fact that uncertainties in the matrix elements are run through
highly nonlinear operations, with the degree of nonlinearity growing with
the number of parameters being optimized.
It would therefore not be surprising if the number of samples needed to
produce a healthy optimization grew with parameter number, especially
considering that, in the extreme limit of a sample size smaller than the
number of optimizable parameters, the LM and SR working equations become
straightforwardly underdetermined and thus ill-defined.
A further complication arises in variance optimization, as the variance
of the energy variance is not finite when employing the usual $|\Psi|^2$
importance sampling function, \cite{Trail2008a,Trail2008b}
although this issue can be resolved by employing other importance
sampling functions. \cite{Robinson2017,Otis2020}

While taking a large enough sample size is in principle sufficient
to overcome these optimizations' compounded statistical uncertainty
issues so long as importance sampling has been handled correctly,
arbitrarily large samples are not always possible in practice and
so alternatives must be pursued.
AD methods have an obvious advantage in this regard, as they avoid
matrix-based working equations entirely and with them the complications 
that come from running statistically uncertain quantities through
highly-nonlinear equations.
While parameter gradients possess an infinite variance when estimated with the 
$|\Psi|^2$ distribution,\cite{Pathak2020} the same strategies for addressing the energy variance's infinite variance, such as employing other importance sampling functions,\cite{Robinson2017,Otis2020} can be straightforwardly applied to these quantities. 
AD methods are thereby able to operate stably with far smaller sample
sizes at each optimization step than either SR or the LM.
In fact, because the damping factors \cite{Schwarz2017}
employed to stabilize the long-term behavior of AD gradually convert
them into stochastic gradient descent (SGD),
they benefit from SGD's guarantee of eventual convergence.
However, this guarantee does not ensure that such convergence occurs
in a manageable number of optimization steps, which in practice means
that the total number of samples needed by AD across all optimization
steps is often larger than for the LM. \cite{Otis2019}
The blocking approach \cite{Zhao2017} and its hybridizations
with AD \cite{Otis2019,Otis2020} exist as a middle ground, reducing
the size of and thus the degree of nonlinearity in the matrix equations
without dropping them entirely.

Now, none of these optimization techniques would be viable without
the ability to efficiently evaluate the necessary derivatives of
the wave function and its local energy at each random sample.
Recent years have seen especially strong progress in this
regard in the area of multi-Slater Jastrow wave functions
with the introduction of fast analytic gradients \cite{Filippi2016,Assaraf2017}
that fully exploit the structure of the table method algorithm.
\cite{clark2011,morales2012}
This and similar approaches have been applied to wave functions
with tens of thousands of determinants, \cite{Dash2019}
excited states, \cite{Flores2019}
orbital-space VMC, \cite{mahajan2020}
and even geometry optimizations. \cite{Dash2019}
These advances in optimization and gradient methods have made
it increasingly possible for VMC to benefit from the recent
progress in sCI methodology, and it is to this combination
to which we now turn.

\subsubsection{Selected Configuration Interaction and VMC}

Although multi-determinant trial functions have long been
used within VMC,\cite{Harrison1985,Umrigar1993,Brown2007,Toulouse2008,Nukala2009,Seth2011,clark2011,morales2012}
the approach has been invigorated in recent years by
connections to progress in sCI methodology.
sCI methodology has improved to the point that,
for molecules with two or three heavy (non-hydrogen)
atoms, there are now many sCI methods available that can
offer affordable, FCI-quality results in large basis sets.
\cite{Evangelista2014,Tubman2016,Holmes2016,Schriber2017,Holmes2017,Sharma2017,Garniron2018,Levine2020,Tubman2020}
Although they are far less expensive than FCI, sCI methods
share its exponential cost-scaling with system size, and so
this exquisite accuracy is, for now at least, not available
in larger molecules.
However, unconverged sCI calculations in such molecules still
contain a wealth of information, and in particular can serve
as an efficient way for identifying the most important
determinants to include in QMC calculations.
\cite{Giner2013,Giner2016,Scemama2018,Scemama2019,Benali2020,Scemama2020,dash2018,Dash2019,Dash2021,Flores2019,Otis2019,Otis2020}
The basic idea is that, although any multi-Slater expansion
extracted from sCI in a large molecule will not be capable of
an exhaustive treatment of electron correlation, it does not
have to be if it can be complemented by Jastrow factors or
diffusion Monte Carlo (DMC).
If the incomplete expansion taken from sCI provides a good
accounting of any strong correlation effects, then wave
function components like Jastrow factors that are better suited
to the details of weak correlation such as cusp effects can
be employed to produce a highly-accurate multi-Slater Jastrow
ansatz.
With the recent advances in optimization methods discussed above,
this approach has made great strides in recent years.

An early ground state application to the oxygen atom\cite{Giner2013} bypassed VMC 
optimization entirely by merely performing DMC with the nodes set by a
pure sCI wave function with no Jastrow term.
A further study on first-row atoms\cite{Giner2016} explored the advantages of reoptimizing 
sCI determinant coefficients together with the Jastrow factor
and more recent molecular applications have employed full VMC optimization of 
wave functions with thousands of determinants to obtain 
VMC geometries in medium-sized molecules. \cite{dash2018,Dash2019,Dash2021} 
In the solid state, systematic reduction of the fixed-node error through 
larger sCI expansions has been demonstrated in diamond.\cite{Benali2020}
Another important success of the combination of sCI and QMC, which we will
add to through the demonstrations below, 
has been the calculation of highly accurate excitation 
energies in molecules. \cite{Dash2019,Flores2019,Cuzzocrea2020,Dash2021,Otis2020}

\subsubsection{Excited States}

Although most uses of QMC methods focus on the ground state, there has also been
a substantial amount of work applying QMC, and especially VMC, to excited states.
In many cases, and especially in earlier applications, researchers have relied
on quantum chemistry methods like DFT and CASSCF to determine
the fermionic part of the wave function, with only the Jastrow factor
optimized within VMC.
\cite{grossman2001,aspuru2004,Tiago2008,bouabcca2009,toulouse2012,zubarev2012}
Although in this Focus Article we will for efficiency's sake advocate a partial
return to this reliance on quantum chemistry in the context of
excited-state-specific CASSCF, we emphasize that these early uses of quantum
chemistry to determine the final orbitals and/or determinant coefficients
typically did not have the luxury of employing excited-state-specific
quantum chemistry methods.

Of course, the optimal orbital shapes and determinant coefficients for an
excited state change upon the introduction of a Jastrow factor, and much
work has focused on relaxing excited states' orbital and CI parameters within VMC.
In many cases, molecular symmetry causes the excited state of interest
to be the lowest energy eigenstate within its irreducible representation,
allowing ground state VMC techniques to be applied.
\cite{zimmerman2009,dubecky2011,dupuy2015}
In other cases, energy minimization with root targeting can be successful,
\cite{zimmerman2009}
but in general this approach is prone to root flipping. \cite{dorando2007,lewin2008}
Multiple approaches have been taken to avoid this issue,
including a recently-introduced penalty method \cite{pathak2021} that,
like similar projector Monte Carlo methods before it, \cite{Blunt2015a}
seeks to optimize an energy eigenstate that is orthogonalized against
lower states.
By far the most widely used approach, however, both to combat root flipping and
in excited state VMC more generally, is state averaging,
\cite{schautz2004a,schautz2004b,Cordova2007,Filippi2009,Valsson2010,Send2011,Filippi2012,Guareschi2013,valsson2013,floris2014,guareschi2014,amovilli2015,Guareschi2016,zulfikri2016,Dash2019,Dash2021}
in which one minimizes the average energy of the ground state and some number
of excited states using a common set of molecular orbitals and (often)
a common Jastrow factor.
Regardless of the optimization approach, an imbalance in the quality of
the descriptions of ground and excited state can lead to poor accuracy
for excitation energies, and so it is becoming increasingly common to
use measures of wave function quality such as the VMC variance \cite{Robinson2017}
or the sCI variance or PT2 correction size \cite{Dash2019,Dash2021}
to ensure this balance is indeed present.

\subsection{Excited-State-Specific Methods}

\subsubsection{Variational Monte Carlo}


Another approach to excited states in VMC is to pursue individual excited states by minimizing an objective function for which they are the unique global minimum.
The existence of such objective functions was recognized at least as early as the late 1960s,
\cite{messmer1969,choi1970}
and they were first connected to VMC theory in 1988. \cite{Umrigar1988}
More recently, this class of objective functions has been explored further in order to improve the ability to target an individual excited state.  \cite{Zhao2016}
The essential strategy behind these VMC-friendly, excited-state-specific objective functions is to exploit the fact that the energy uncertainty $\sigma$ (or, equivalently, the energy variance $\sigma^2$) is zero if and only if the wave function is a Hamiltonian eigenstate.
Thus, each non-degenerate excited state exists as a local minimum of the energy variance.
With some manipulation, objective functions related to the energy variance
\cite{Umrigar1988,Zhao2016,Shea2017,Cuzzocrea2020}
can be built such that a user-controlled energy value $\omega$ can be used to select which excited state will be the global minimum and thus targeted by the optimization.
Interestingly, the simplest versions of these objective functions break size-consistency if the targeting energy $\omega$ is held fixed, but in practice it can be automatically adjusted on the fly so as to turn these approaches into excited-state-specific variance minimization methods.  \cite{Shea2017}

It is important to note that care must be taken in the design of the numerical methods used in the nonlinear minimization of these objective functions to ensure they are stable.
As has been recently reported, some numerical methods choices can lead to unstable, failed optimizations.
\cite{Cuzzocrea2020}
Although a technical discussion of this issue lies outside
the scope of this Focus Article, we will say that our own investigations
indicate that such instabilities in excited-state-specific variance optimization
arise from the way in which statistical uncertainty interacts with
optimization step control measures such as
step damping and trust radius methods.
Some classes of step control appear to be ineffective in this context,
while others are well-suited for dealing with the ways in which
statistical uncertainty affects parameter update steps.
We will address this topic in detail in an upcoming publication.
For this Focus Article, let us simply stress
that the step control measures built into the hybrid optimization
method discussed below produced stable optimizations in all
the states investigated for this article.
Beyond the design of the numerical minimization method, it is also important to use
importance sampling functions that avoid the infinite variance-of-the-variance issue
\cite{Trail2008a,Trail2008b,Robinson2017}
that, if not addressed, prevents accurate estimations of these objective functions.

Although the application of excited-state-specific VMC to real molecular and materials systems is still in its early days, a number of promising results can already be found in the literature.
These include modeling doubly excited states with both geminal power wave functions \cite{Zhao2016}
and multi-Slater Jastrow expansions. \cite{Otis2020}
Other applications involving multi-Slater Jastrow wave functions
include charge transfer states, \cite{Flores2019}
core excitations, \cite{Garner2020a}
and the band edge optical states in solids. \cite{zhao2019a}
These objective functions have also supported the development of a variation-after-response wave function ansatz
\cite{neuscamman2016,Blunt2017a}
and its use as a starting point for excited-state diffusion Monte Carlo.
\cite{blunt2018}
These examples show that excited-state-specific VMC has promise in its own right, but, as we will argue in this Focus Article, the opportunities that arise when it is combined with excited-state-specific quantum chemistry are especially compelling.

\subsubsection{Quantum Chemistry}

Excited-state-specific methods in quantum chemistry also have a long history
and have benefited from a number of recent developments.
Perhaps the longest-standing approach is $\Delta$-self-consistent-field
\cite{bagus1965scf,Pitzer1976scf,argen1991xray,Gill2009dscf}
($\Delta$SCF), in which one seeks to converge the optimization of a
single-determinant wave function to a non-ground-state energy stationary point,
typically by solving for higher roots of the Roothaan equation.
The efficacy of this approach was substantially improved by the introduction
of maximum overlap methods, \cite{gilbert2008self,barca2018simple}
and can be made more robust still by using quantum-chemistry-friendly
excited state variational principles based on the norm of the energy gradient.
\cite{shea2020,Hait2020a}
Similarly, it is also possible to use energy-variance-based variational principles
in quantum chemistry, which has facilitated the development of the
$\sigma$-SCF method for excited states. \cite{ye2017,ye2019}
Other excited-state-specific approaches that center on mean-field ideas
include constrained DFT, \cite{VanVoorhis2005cdft}
square-gradient-minimization DFT \cite{Hait2020a,hait2020b,hait2020c,Hait2021}
(SGM-DFT), and excited state mean field theory
\cite{shea2018,zhao2019b,zhao2020,hardikar2020,garner2020b,shea2020}
and its corresponding second-order perturbation theory (ESMP2). \cite{shea2018,clune2020}
Very recently, more robust methods for excited-state-specific optimizations
of the multi-reference complete active space self-consistent field (CASSCF)
method have also been introduced. \cite{Tran2019,Tran2020,Hanscam2021}
Together, these advances allow post-excitation orbital relaxations to be treated
on an equal footing with ground state approaches, which contrasts with
state-averaging methods (where multiple states must come to
a compromise over the orbital shapes),
EOM-CC (where orbital relaxations are approximated
by linear response),
and TDDFT (where the adiabatic approximation prevents orbital relaxations
for all electrons other than the one being excited).

The recent advances in excited-state-specific quantum chemistry are already
offering new capabilities in challenging excited state scenarios.
Both SGM-DFT \cite{hait2020b} and ESMP2 \cite{garner2020b} now offer
more accurate descriptions of valence relaxations in the presence of a
core hole for 2nd-row K-edge states than EOM-CC, and the former has been
successfully extended to core excitations in radical species. \cite{hait2020c}
Although there is only a limited amount of data in the literature so far,
ESMP2 has shown significant promise for charge transfer excitations.
\cite{clune2020}
For double excitations dominated by a single determinant, both SGM-DFT
\cite{Hait2020a} and $\sigma$-SCF \cite{ye2017} show considerable promise.
Beyond vertical excitation energies, state-specific CASSCF (SS-CASSCF)
has proven capable of making qualitative improvements to potential
energy surfaces as compared to state-averaging \cite{Tran2020}
and in some cases offers smooth surfaces where state-averaging
creates artificial discontinuities. \cite{Tran2019}
As we will now discuss and demonstrate through examples,
SS-CASSCF is an excellent example of a quantum chemistry method
whose combination with excited-state-specific VMC offers
significant opportunities.

\subsubsection{A Promising Intersection}

In the same way that ground state quantum chemistry methods
are often used to supply good initial wave functions to ground
state quantum Monte Carlo methods, excited-state-specific
methods in quantum chemistry and quantum Monte Carlo are
maturing to the point that a similar partnership holds
great promise for modeling excited states.
The remainder of this Focus Article will argue this point
while also providing demonstrations of what
is already possible with a particular combination of
excited-state-specific quantum chemistry and
quantum Monte Carlo.
This combination will focus on SS-CASSCF,
sCI, and excited state variance minimization in VMC, aiming to provide
reliable accuracy across singly excited, doubly excited,
and charge transfer excitations in contexts beyond the
reach of DMRG, FCI-QMC, and sCI-PT2.
However, this combination is just one example
of the opportunities at this intersection, and it does
not touch on major areas like projector Monte Carlo
and excited-state-specific DFT.
We therefore hope that it can help motivate
further explorations at this intersection,
from which we have every expectation that even more
promising methods will arise.

The ability of SS-CASSCF to deliver excellent excited state
orbital shapes allows it, in conjunction with sCI and
state-specific variance minimization, to partner with VMC
to deliver the orbital relaxation, strong correlation,
and weak correlation treatments necessary for reliable
accuracy in a wide range of excited states.
Although it is increasingly possible to optimize orbital
shapes together with large determinant expansions and
Jastrow factors in VMC, \cite{Filippi2016,Assaraf2017}
our experience has been that VMC orbital optimization
remains more expensive and challenging than VMC
optimizations in which the orbitals are held fixed.
Hoping that SS-CASSCF has already done a good job at
providing any necessary post-excitation orbital relaxation,
we therefore hold our orbitals fixed after the
SS-CASSCF calculation is complete.
Although this will certainly limit the absolute
accuracy of the individual states, we will see that it
does not prevent the overall approach from delivering
reliably high accuracy for excitation energies.
We expect that this success for SS-CASSCF orbitals
is due in part to our post-CASSCF use of sCI
in a greatly expanded active space, which, among other
things, can pick up determinants that are singly-excited
out of the smaller SS-CASSCF active space and thus
approximate further orbital relaxations that may be
appropriate in light of correlation effects.
It is important to note that this sCI is not run
to convergence as it need only identify the most important
determinants for use in VMC, allowing the extended active
space to reach beyond what sCI methods could tackle on
their own.
Like many combinations of sCI and QMC,
\cite{Caffarel2016,Scemama2014,Scemama2018,Scemama2019,Dash2019,Dash2021,Benali2020}
this approach benefits from the relative sparsity
of most low-lying electronic states in Fock space
and the ability of Jastrow factors
\cite{Umrigar1988,Drummond2004,LopezRios2012,Luchow2015,Huang1997,Casula2003,Casula2004,Sterpone2008,Beaudet2008,Marchi2009,Zen2015,Goetz2017,Goetz2019}
to address much of the remaining weak correlation
treatment that exhaustive determinant expansions
are inefficient for.

Since our orbital shapes are already state-specific,
we rely on a recently developed approach to
state-specific variance minimization \cite{Otis2020}
to further optimize the CI coefficients in tandem
with the Jastrow parameters.
Again, it is possible to also relax the orbitals at this point,
but we have found that this does not offer much advantage in
terms of accurate excitation energies relative to the
starting SS-CASSCF orbitals.
Finally, since these multi-Slater Jastrow wave functions
will still not be exhaustively accurate in larger molecules
and basis sets,
we balance the quality of the ground and excited state
through VMC variance matching.
\cite{Robinson2017,Flores2019,Garner2020a,Otis2020}
Recent work suggests that it may be even more effective to perform
quality matching at the sCI stage, \cite{Dash2019,Dash2021}
but for the demonstrations in this article we have employed
VMC variance matching for simplicity.
As a final note on the method we will demonstrate, it is worth
paying attention to the relatively small determinant expansions
that are employed.
While it is now possible to use tens and even hundreds
of thousands of determinants in VMC, the more determinants that
are employed, the more limited the approach will be in the system
sizes it can treat.
Indeed, the ability of Jastrow factors and variance matching
to mitigate the shortcomings of modest determinant expansions
makes it easier to deploy the method in situations
(e.g.\ the charge transfer state of benzonitrile in aug-cc-pVTZ)
that are beyond the current reach of methods like sCI-PT2
and FCIQMC.
To emphasize this point, let us stress that none of the
calculations presented below required parallelization beyond
ten 40-core computing nodes, which, at least by the standards
of QMC, is not a huge computational investment.

This combination of quantum chemistry with VMC offers a number of key advantages.
The explicitly multi-reference character is essential when treating
difficult double excitations and in states with significant mixtures
of singly- and doubly-excited character.
By avoiding state averaging, charge transfer states are provided with fully
appropriate post-excitation orbital relaxations.
For weak correlation effects, the use of Jastrow factors and a modest but sCI-chosen
configuration interaction expansion appears to be sufficient for the
error cancellation balance of variance matching to do its job and deliver
0.1 eV accuracies in all systems tested in which a reliable theoretical benchmark exists.
This interplay between variance matching, Jastrow factors, and the limited CI
expansion is crucial, as it allows us to push into larger molecules and
basis sets than could be reached by relying on determinant expansions alone.
Compared to widely used quantum chemistry methods like TDDFT, EOM-CC, and CASPT2,
the approach is noteworthy in its ability to offer 0.1 eV accuracy 
across singly excited, doubly excited, and charge transfer states, which
none of these quantum chemistry methods can deliver across all of these
excitation classes.

It is important to emphasize that the approximations
employed in this approach are quite different than those used in EOM-CC,
which allows us to validate coupled cluster
results in some circumstances where it is not otherwise obvious
(e.g.\ due to partial doubly excited character) how accurate EOM-CC should be.
When these methods agree about a challenging excitation energy,
such as in the $3^1A'$ state of uracil, the fact that they rely on very
different assumptions and approximations increases our confidence in the
accuracy of the prediction.
We therefore hope that, in addition to its longstanding role in
providing ground state reference data in both molecular
\cite{Austin2012} and solid-state \cite{Foulkes2001} 
settings, quantum Monte Carlo can, in tandem with excited-state-specific
quantum chemistry and coupled cluster theory, be increasingly helpful
in this regard in challenging molecular excited states.

\section{Theory}

\subsection{Excited State-Specific Variational Monte Carlo}

The application of VMC to excited states is an active area
of research with multiple optimization objectives
that can be employed, including 
both state-averaged energy minimization ($E^{SA}$)
\cite{Cordova2007,Filippi2009,Send2011,Guareschi2013,Guareschi2016,Cuzzocrea2020,Dash2021} 
and state-specific variance-based minimization
($\Omega$ and $W$).
\cite{Umrigar1988,Zhao2016,Shea2017,Flores2019,Garner2020a,Otis2020}

\begin{equation}
\label{eqn:targetAbove}
    \Omega (\Psi) = \frac{\Braket{\Psi | (\omega - H) | \Psi}}{\Braket{\Psi | (\omega - H)^2 | \Psi}} = \frac{\omega - E}{(\omega - E)^2 + \sigma^2}
\end{equation}

\begin{equation}
\label{eqn:targetClosest}
    W (\Psi) = \Braket{\Psi | (\omega - H)^2 | \Psi} = (\omega - E)^2 + \sigma^2
\end{equation}

\begin{equation}
\label{eqn:stateAverageE}
    E^{SA} = \sum_I w_I\frac{\Braket{\Psi^I | H | \Psi^I}}{\Braket{\Psi^I  | \Psi^I}}
\end{equation}
For the demonstrations in this article, we employ the
state-specific objective function \cite{Zhao2016} $\Omega$,
which targets the lowest state above $\omega$ in energy when minimized.
$\Omega$ and $W$ are nonlinear functions of the energy $E$ and variance $\sigma^2$, which 
leads to a failure of size-consistency.\cite{Shea2017}
This can be verified by considering subsystems $A$ and $B$ with a product factorizable ansatz. 
In this case, the energy and variance will be additive ($E = E_A + E_B$ and $\sigma^2 = \sigma^2_A + \sigma^2_B$), but $\Omega$ and $W$ will not, due to their nonlinearity.
To achieve size-consistent results, the input $\omega$ must
be updated to $E - \sigma$, the difference between the energy and
its standard deviation, which transforms the optimization of
$\Omega$ into state-specific variance minimization. \cite{Shea2017}
There are multiple approaches to varying $\omega$, but in this work we
simply conduct multiple fixed-$\omega$ calculations and change the value
of $\omega$ between each one until self-consistency between it and
$E -\sigma$ is reached.
In practice, only a few such fixed-$\omega$ optimizations are needed
for any one state.

In contrast to the deterministic methods that are predominantly used
in quantum chemistry, QMC makes stochastic estimates of the
energy and all related quantities such as gradients and second derivatives.
The quality of these estimates can depend on the probability distribution
from which samples are drawn, and practitioners have useful flexibility
in the choice of this distribution.
While
$\rho (\mathbf{R}) = \frac{\Psi(\mathbf{R})^2}{\int d \mathbf{R} \Psi (\mathbf{R})^2}$
is a common choice and has a zero variance property \cite{Assaraf1999}
for the energy when $\Psi$ is an  exact eigenstate, it suffers from
an infinite variance problem with approximate $\Psi$
in the  estimation of other key quantities, particularly the variance
\cite{Trail2008a,Trail2008b,Robinson2017}
and some wave function derivatives.
To address this issue, regularization schemes,
\cite{Pathak2020,vanRhijn2022} estimator modifications,
\cite{Assaraf1999,Assaraf2000,Assaraf2003}
and other importance sampling functions
\cite{Attaccalite2008,Trail2008a,Trail2008b,Robinson2017,Flores2019,Otis2020}
have all been variously employed by the QMC community.
In this study, we use the same importance sampling function
\begin{equation}
\label{eqn:guiding}
    |\Phi|^2 = |\Psi|^2 + c_1\sum_{i} |\Psi^i|^2 + c_2\sum_{j}|\Psi^j|^2 + c_3\sum_{k}|\Psi^k|^2
\end{equation}
that was previously used and more fully discussed in recent work.\cite{Otis2020}
The coefficients $c_1$,$c_2$,$c_3$ weight the sums of squares of wave function parameter derivatives,
divided between Jastrow, CI, and orbital parameters respectively.
We have found that setting $(c_1,c_2,c_3) = (0.0, 0.0001, 0.0)$ is a reasonably effective choice 
for the molecules we consider.

\subsection{Hybrid Optimization}

The stochastic nature of QMC increases the difficulty of its wave function ansatz optimization 
compared to the analogous problem encountered in deterministic methods.
Optimization has been a long-running challenge in QMC method development for many years 
with a variety of algorithms being introduced. 
These include the linear method\cite{Nightingale2001,Umrigar2007,Toulouse2007,Toulouse2008} and 
accelerated descent\cite{Schwarz2017,Sabzevari2018,Luo2018,Mahajan2019,Otis2019} approaches.
Neither of these classes of methods is fully satisfactory, with the LM possessing
a tendency for poor parameter updates due to stochastic step 
uncertainty with large numbers of variables and AD suffering from slow 
convergence to optimal parameter values.

We have recently developed a hybrid combination of the blocked version of the 
LM\cite{Zhao2017} with AD, first for ground state energy minimization\cite{Otis2019} and 
then for variance-based excited state optimization.\cite{Otis2020}
In this approach, alternating between sections of descent optimization and the blocked LM 
enables more efficient convergence to the minimum than using either method on its own.
The blocked LM is able to move parameters to the vicinity of the minimum more swiftly than 
descent, but descent is able to correct for any poor parameter updates made by the blocked LM
due to step uncertainty, which makes it useful for converging more tightly.
Throughout all our optimizations, we alternate between 100 iterations of AD and 3 steps of blocked
LM and refer to one set of both as a macro-iteration.
A typical optimization consists of a small number of macro-iterations followed by 1100 iterations 
of AD with final energy, target function, and variance averages taken over the final 500 
iterations.
These averages and the excited state optimizations are then used in our variance matching procedure, 
discussed below, to obtain a final excitation energy result.
More extensive discussions of our hybrid method can be found in earlier 
papers\cite{Otis2019, Otis2020} and in this Focus Article, we use the same combination of 
Nesterov momentum with RMSprop and the same hyperparameter choices as before.\cite{Otis2020}

In this study, we improve the efficacy of this hybrid approach further
by filtering which parameters are involved in the blocked LM steps
at all by the statistical significance of the target function's
parameter derivatives, an approach that has been used successfully
in other optimization contexts before. \cite{Nakano2020}
We employ the criterion
\begin{equation}
\label{eqn:filterParam}
   \abs{\frac{<\frac{\partial \Omega}{\partial p} >}{\sigma_{\frac{\partial \Omega}{\partial p}}}} > 1
\end{equation}
to determine which parameters will be involved in the blocked LM steps.
Thus, the mean value of the parameter derivative must be at least as large as the 
standard deviation of that derivative estimate in order for the parameter to be 
included for the blocked LM update on that iteration.
The number and identity of the parameters that are turned off
by this filtration stage will in practice vary from LM step to LM step,
but we have found that it is typical for roughly a third of the 
parameter set to be filtered out in any given blocked LM iteration
in our optimizations.

We also note that the filtered parameter space is further reduced by the blocked LM 
algorithm,\cite{Zhao2017} which divides the parameter set into blocks for LM-style diagonalizations. 
A limited number of eigenvectors are retained from each block to form the space for a second 
diagonalization that determines the final parameter update.
In our optimizations, we have found that dividing the parameters that
survive the filtration into 5 blocks and  retaining 30 parameter directions
from each block for the second stage of the blocked LM to be effective.
The use of the initial filtration stage allows for an automated
identification of a smaller variable space that the blocked LM can
thereby treat with reduced step uncertainty.
The AD sections of the hybrid method continue to optimize all of
the variational parameters.
This refinement of the hybrid method enhances the 
blocked LM's ability to make significant updates to the most
important parameters while allowing AD to handle the remaining
finer-grained parameter adjustments.

\subsection{Wave Functions}

The wave function ansatzes we use in our VMC optimizations are constructed from a preceding 
set of quantum chemistry calculations.
In particular, we obtain molecular orbitals from a recently developed SS-CASSCF
approach, \cite{Tran2019,Tran2020}  which uses a density-matrix-based
scheme to track individual states during CASSCF's CI and orbital
relaxation optimization steps.
This tracking approach overcomes root-flipping and obtains orbitals
tailored specifically for the targeted excited state.
We then employ sCI, specifically heatbath CI, \cite{Holmes2016,Sharma2017}
to generate a set of the most important determinants for the state
from an expanded active space within the now-state-specific orbital basis.
We emphasize that converging this sCI is unnecessary and that in
practice we often need only the first few hundred to few thousand
most important determinants for use within VMC.
With the addition of spline-based one- and two-body Jastrow factors,
we arrive at the form of the multi-Slater Jastrow (MSJ) ansatz used
for all the molecules considered in this work.

\begin{equation}
\label{eqn:psi}
    \Psi = \psi_{MS} \psi_J 
\end{equation}
\begin{equation}
\label{eqn:psiMS}
    \psi_{MS} = \sum_{i=0}^{N_D} c_i D_i
\end{equation}
\begin{equation}
\label{eqn:psiJ}
    \psi_J = \exp{\sum_i \sum_j \chi_{ij}(|r_i - R_j|) + \sum_k \sum_{l>k} u_{kl} (|r_k - r_l|)}
\end{equation}
We construct the Jastrow factor $\psi_J$ with splines for the functions $\chi_k$ and $u_{kl}$.\cite{Kim2018}
As we will show in our results, only limited numbers of the determinants $D_i$ are 
needed to obtain highly accurate excitation energies in many molecules.

In all our VMC results, only the Jastrow factors and the determinant coefficients $c_i$ are 
optimized by our hybrid method, while the orbitals are left at their SS-CASSCF shapes.
While recent developments\cite{Filippi2016,Assaraf2017} with the table method have made 
orbital optimization at the VMC level more efficient, it remains more computationally demanding than the Jastrow and CI optimization, 
and significantly increases the difficulty of our optimization problem.
Our decision to use SS-CASSCF orbital shapes and avoid any VMC orbital optimization enables 
us to obtain significant computational savings while retaining the benefits of state-specificity,
particularly in cases with important orbital relaxation effects such as charge transfer.
As our results show, leaving the orbitals at their SS-CASSCF shapes does not prevent us from obtaining accurate results, suggesting that the lion's share of orbital relaxation is present already in the active space picture.

\subsection{Variance Matching}
\begin{figure}[H]

 \includegraphics[width=\columnwidth]{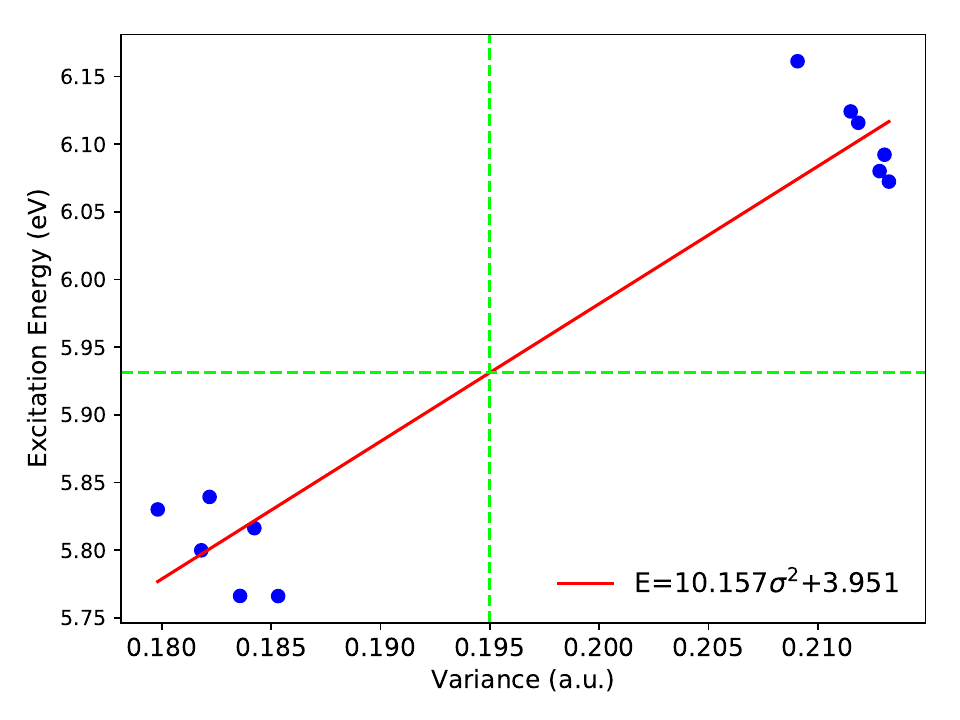}
 \caption{Example of the linear interpolation used in variance matching. The points are  
 variances and corresponding excitation energies for individual blocked LM iterations in ground state optimizations of 100 and 500 determinants. Each set of blocked LM points is taken from the last two macro-iterations of the converged hybrid method optimization. The excited state variance and the matched excitation energy value are 
 marked by the green dashed lines. As shown in Table \ref{tab:small_doubles_data} below, the resulting excitation energy prediction of 5.93 eV is in good agreement with literature benchmarks.
 }
 \label{fig:var_match_plot}

 \end{figure}

Another key part of obtaining a high level of accuracy with compact wave functions is ensuring a 
balanced treatment of both ground end excited state in order to benefit from cancellation of errors.
Past work\cite{Robinson2017,Flores2019,Garner2020a,Otis2020} has shown that matching the variance between ground and excited state improves 
excitation energy predictions, and in our demonstrations below we adopt a recently developed 
interpolation procedure\cite{Garner2020a} and in one case obtain an explicit variance match.
In our interpolations, we perform ground state optimizations at multiple determinant expansion lengths to straddle the 
excited state variance and linearly interpolate to determine the excitation energy for a 
variance match.
An example is shown in Figure \ref{fig:var_match_plot} for the case of the carbon trimer and 
further discussion of the procedure can found in the original publication.\cite{Garner2020a}

\section{Results}

\subsection{Computational Protocol and Technical Details}
All our VMC calculations used a development version of QMCPACK.\cite{Kim2018,Kent2020}
In all cases, we used the pseudopotentials and associated basis sets developed by Mitas 
and coworkers.\cite{Bennett2017}
Molecular geometries and theoretical best estimates (TBEs) were taken from
literature benchmark sets, \cite{Loos2018,Loos2019,Veril2021,Schreiber2008,Kozma2020}
with the molecular coordinates also listed in Appendix A.
State-specific molecular orbitals were generated by an implementation
of SS-CASSCF\cite{Tran2019,Tran2020} in a 
development version of PySCF.\cite{Sun2018}
We generated sCI expansions in expanded active spaces with Dice,
\cite{Holmes2016,Sharma2017} iterating until at least a million determinants
had been selected and then taking those with the largest coefficients into our 
VMC wave function.
The particular active space and basis set choices for different molecules are
given in the following sections for each set of excited states.
For the charge transfer states we study, we have performed CASPT2 and coupled cluster calculations in 
Molpro\cite{MOLPRO_brief} and GAMESS\cite{GAMESS2020} for comparisons against our VMC results.

Once our ansatzes were generated, the CI coefficients were optimized within VMC simultaneously with one- and two-body Jastrow factors.
The Jastrow factor splines each consisted of 10 coefficients defining the function within a distance 
of 10 bohr.
With two exceptions, the initial value of the Jastrow coefficients was set to zero in all cases and 
CI coefficients were begun at their values from sCI. 
In the case of nitrosomethane, the final 1000 determinant excited state optimization began 
from the result of a previous optimization of the Jastrow and 500 determinants, with the next 500 
most important determinants from sCI added with initial coefficients of zero.
For benzonitrile, the final ground and excited state optimizations began from 
the respective optimized wave functions obtained with initial guesses for $\omega$.

\subsection{Valence Single Excitations: Thioformaldehyde, Methanimine, and Ketene}
In testing our methodology, we begin by considering valence single excitations, specifically
the $  {}^1 A_2 (n \rightarrow \pi^*)$ excited state in 
thioformaldehyde, the $  {}^1 A^{''} (n \rightarrow \pi^*)$ state in methanimine, and the $  {}^1 A_2 (\pi \rightarrow \pi^*)$ state in ketene. 
For these three excited states, sCI and high level coupled cluster 
results\cite{Loos2018} from Loos and coworkers 
are reliable references for assessing the accuracy of our approach.
We generated our SS-CASSCF orbitals using a (12e,10o) active space for thioformaldehyde, a (12e,12o) 
space for methanimine and a (8e,8o) space for ketene, with pseudopotentials and cc-pVTZ basis sets 
for all molecules.\cite{Bennett2017}
The subsequent sCI calculations in Dice\cite{Holmes2016,Sharma2017} correlated 12 electrons in 17 
orbitals for thioformaldehyde, 12 electrons in 18 orbitals for methanimine, and 16 electrons in 22 
orbitals for ketene.

\begin{figure}[H]

\includegraphics[width=\columnwidth]{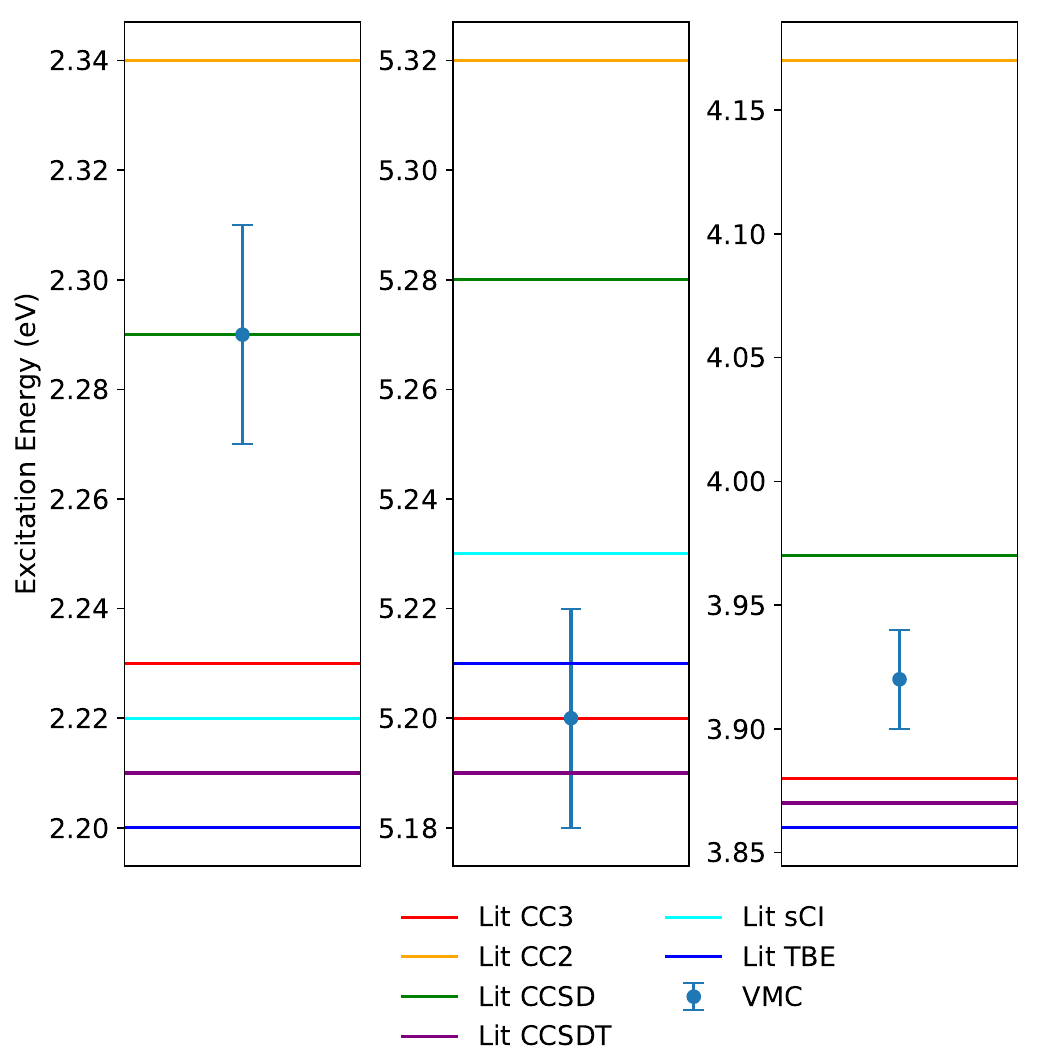}
\caption{From left to right, excitation energies in 
thioformaldehyde, methanimine, and ketene for VMC and quantum chemistry approaches. See also Table \ref{tab:full_singles_data}. Reference values for coupled cluster and selected CI are taken from the work of Loos and coworkers.\cite{Loos2018}}
\label{fig:full_single_results}

\end{figure}

\begin{table}[H]

\caption{Excitation energies for thioformaldehyde, methanimine, and ketene. The VMC uncertainties on the final digits are given in parentheses. The included literature values use an aug-cc-pVTZ basis set for sCI  and coupled cluster, and the literature TBEs are obtained from applying a basis set correction using CC3 to the sCI result.\cite{Loos2018}}
\begin{tabular}{lSSS}
Method & \multicolumn{1}{l}{Thioformaldehyde}  & \multicolumn{1}{l}{Methanimine}  & \multicolumn{1}{l}{Ketene} \\ \hline

VMC    & 2.29(2)        & 5.20(2) & 3.92(2)\\
CC2    & 2.34        &   5.32       & 4.17\\
CCSD    & 2.29        &  5.28       & 3.97\\
CC3  & 2.23 &           5.20        & 3.88\\ 
CCSDT  & 2.21 &         5.19        & 3.87\\
sCI  & 2.22 &            5.23       & 3.86\\
Lit TBE  & 2.20 &       5.21        & 3.86\\
\end{tabular}

\label{tab:full_singles_data}
\end{table}

Our results for this trio of small molecules, depicted visually in Figure 
\ref{fig:full_single_results} with precise values in Table 
\ref{tab:full_singles_data}, show that our VMC methodology is 
highly accurate.
In all cases, our final excitation energies are within 0.1 eV of the theoretical best estimates 
determined by Loos and coworkers from basis set corrected sCI calculations.
We also emphasize that our results require only relatively simple wave function ansatzes.
For thioformaldehyde, we used 300 determinants for the excited state and the variance matching interpolation used 700 and 1000 determinant calculations for the ground state. Methanimine used 1000 determinants for the excited state while the ground state calculations used 100 and 300. In the case of ketene, the excited state used 700 determinants and 50 and 500 determinants were used for the ground state calculations.
Thioformaldehyde was somewhat unusual with the ground state having the higher variance for fewer 
determinants so longer expansions were optimized to straddle the excited state variance for performing the matching interpolation.
The VMC wave functions used in these systems are significantly less complex than those obtainable 
by coupled cluster and the absolute energies from VMC are comparable to 
those of CCSD.
The excitation energy differences from VMC are also most comparable to CCSD in terms of 
accuracy, as CC3 and CCSDT are within one or two hundredths of an eV of the sCI-based TBEs.
While our approach does not obtain the exquisite accuracy of higher level coupled cluster in these cases, these
results are a reassuring check that our approach agrees 
with very well established quantum chemistry treatments of this generally easier class of 
excited states, and we now turn our attention to more challenging categories.

\subsection{Double Excitations: Carbon Trimer, Nitrosomethane, Hexatriene, and Benzene}

We next address double excitations, which are a far more challenging class of excited states
for coupled cluster.
Past work\cite{Otis2020} applying VMC to double excitations showed a high level of accuracy and we
again draw from the wealth of benchmarking data\cite{Loos2019} from Loos and 
coworkers, comparing our SS-CASSCF/sCI/VMC methodology to their quantum
chemistry values.
We first consider two small systems, the carbon trimer and nitrosomethane, where highly reliable 
excitation energies can still be established by sCI, before studying hexatriene and benzene, which are
considerably larger cases where less definitive methods must be used.
For our ansatz construction, SS-CASSCF orbitals were obtained from a (12e,12o) space for the carbon 
trimer, a (12e,9o) space for nitrosomethane, and (6e,6o) spaces for hexatriene and benzene.
At the sCI stage, these spaces were expanded to (12e,18o), (18e,22o), (32e,46o), and 
(30e,46o) respectively.
For all four molecules, we have used cc-pVTZ basis sets and pseudopotentials.\cite{Bennett2017}

\begin{figure}[H]

\includegraphics[width=\columnwidth]{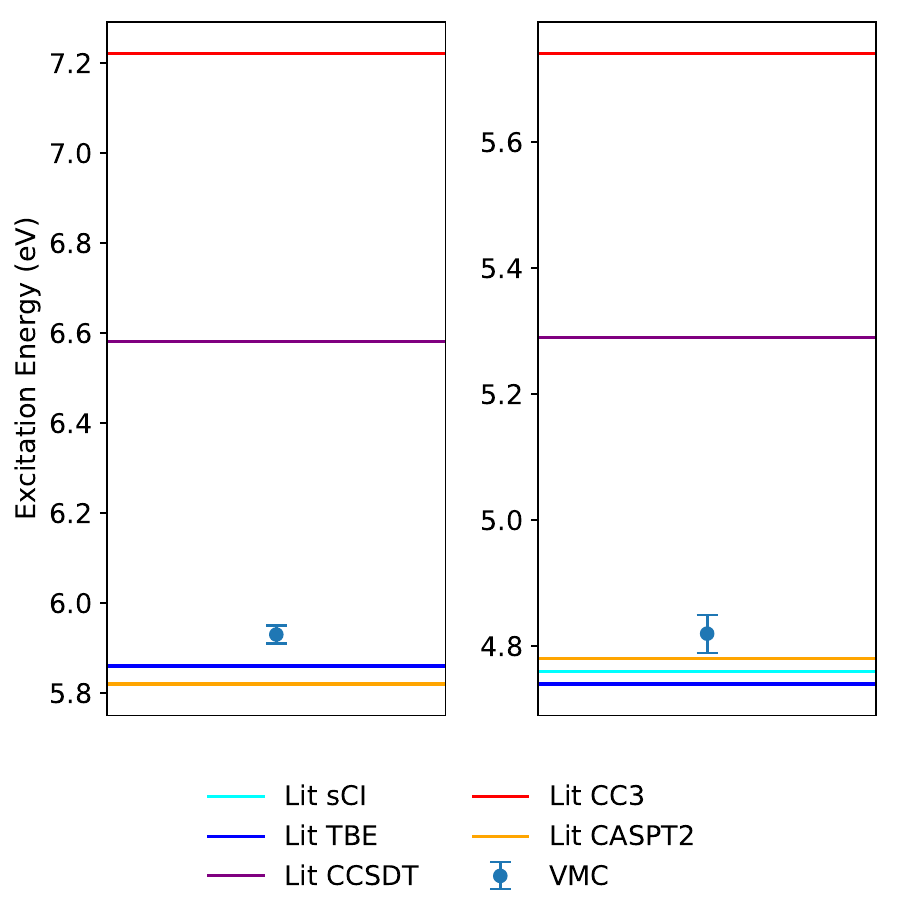}
\caption{Excitation energies in 
the carbon trimer and nitrosomethane in the left and right panels respectively. See also Table \ref{tab:small_doubles_data}. For the carbon trimer, VMC used 500 determinants for the excited state and interpolated between 100 and 500 for the ground state. Nitrosomethane used 1000 determinants for the excited state and interpolated between 10 and 300 for the ground state. Reference values for coupled cluster, and selected CI are taken from the work of Loos and coworkers.\cite{Loos2019}}
\label{fig:small_doubles_results}

\end{figure}

Concentrating first on the two smaller systems, we see in Figure \ref{fig:small_doubles_results} and Table \ref{tab:small_doubles_data} that our VMC approach is again highly accurate and within 0.1 eV 
of the TBEs obtained from sCI.
Both states can be classified as pure double excitations based on the low percentage of $T_1$
amplitudes in CC3\cite{Loos2019} and we see that CC3 makes substantial 
errors of an eV or worse for this type of doubly excited state.
Shouldering the expense of full triples with CCSDT only manages to eliminate about half of CC3's errors.
In contrast, CASPT2's multi-reference nature makes it far more successful in this situation and it 
achieves comparable accuracy to our VMC results in both molecules.

\begin{table}[H]

\caption{Excitation energies for 
the carbon trimer and nitrosomethane. The VMC uncertainties on the final digits are given in parentheses. The included literature\cite{Loos2019} values use an aug-cc-pVQZ basis set for the carbon trimer and the CC3 and CASPT2 values for nitrosomethane. The literature nitrosomethane CCSDT and sCI results used an aug-cc-pVTZ basis set.}

\begin{tabular}{lSS}
Method & \multicolumn{1}{l}{Carbon Trimer}  & \multicolumn{1}{l}{Nitrosomethane}  \\ \hline

VMC    & 5.93(2)        & 4.82(3)    \\
CC3  & 7.22 &         5.74       \\
CCSDT  & 6.58 &         5.29       \\
CASPT2  & 5.82 &        4.78        \\
sCI    & 5.86 &      4.76         \\
Lit TBE     & 5.86 & 4.74
\end{tabular}

\label{tab:small_doubles_data}
\end{table}

\begin{figure}[H]

\includegraphics[width=\columnwidth]{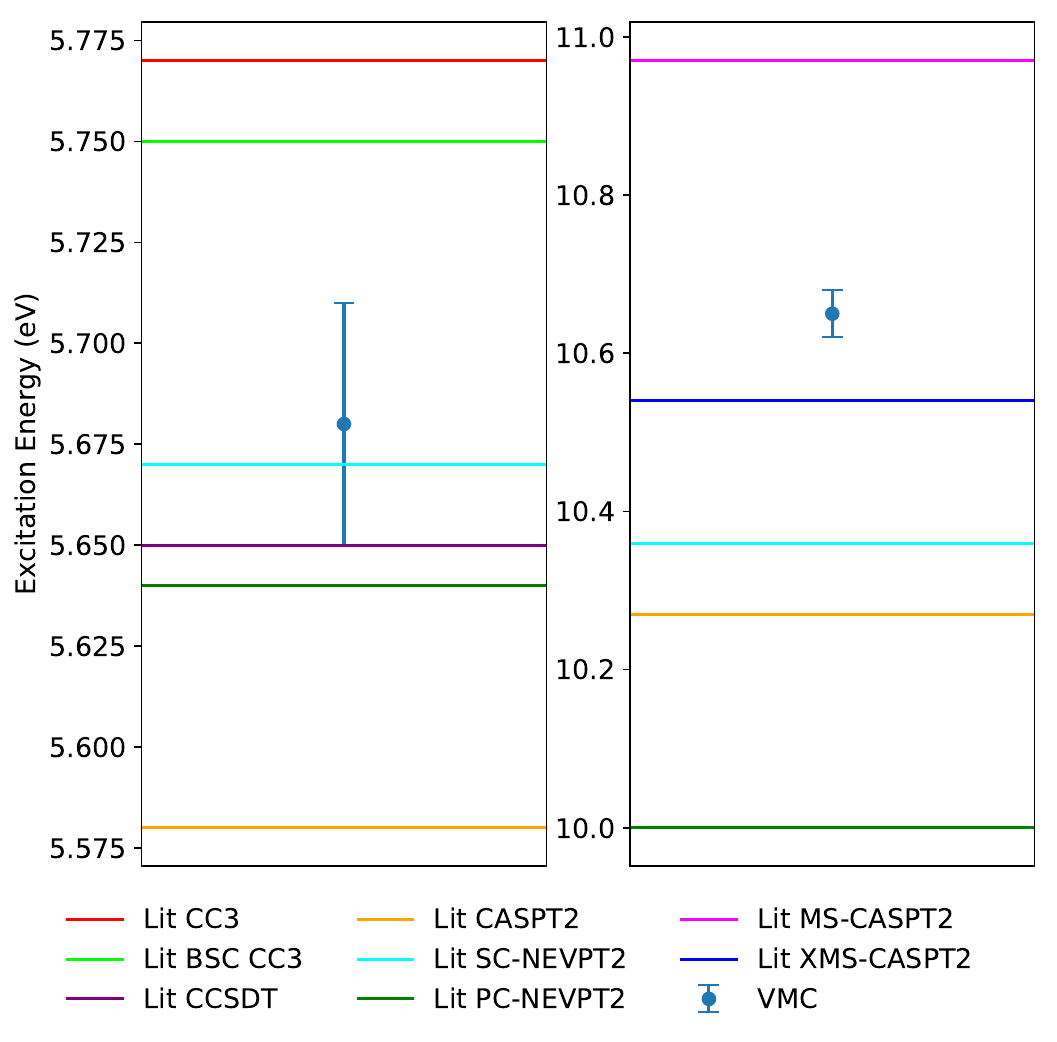}
\caption{Excitation energies in hexatriene and benzene in the left and right panels respectively. See also Table \ref{tab:large_doubles_data}. In both molecules, VMC used 1000 determinants for the excited state and interpolated between 5 and 500 for the ground state.} Reference values for coupled cluster, CASPT2 and NEVPT2 are taken from the work of Loos and coworkers.\cite{Loos2019}
\label{fig:large_doubles_results}

\end{figure}

\begin{table}[H]

\caption{Excitation energies for 
hexatriene and benzene. The VMC uncertainties on the final digits are given in parentheses. The included literature\cite{Loos2019} values use an aug-cc-pVDZ basis set for CC3 and CCSDT in hexatriene and an aug-cc-pVQZ basis set for all varieties of CASPT2 and NEVPT2, except for PC-NEVPT2 in benzene, which used aug-cc-pVTZ.}

\begin{tabular}{lSS}
Method & \multicolumn{1}{l}{Hexatriene}  & \multicolumn{1}{l}{Benzene}  \\ \hline

VMC    & 5.68(3)        & 10.65(3)     \\
CC3  & 5.77         &                \\
BSC CC3  & 5.75         &                \\
CCSDT  & 5.65       &                \\
CASPT2  & 5.58      &        10.27       \\
PC-NEVPT2 & 5.64    &        10.00     \\
SC-NEVPT2  & 5.67   &        10.36       \\
MS-CASPT2  &        &          10.97     \\
XMS-CASPT2  &       &         10.54      \\
\end{tabular}

\label{tab:large_doubles_data}
\end{table}

For molecules at the scale of hexatriene and benzene, high accuracy excitation energies from sCI are 
now unaffordable and various flavors of coupled cluster and multi-reference perturbation theory are 
left as the main quantum chemistry tools.
This leaves space for our VMC approach to provide reliable excitation energies that can help 
determine whichever other methods are most accurate in larger systems.
For the hexatriene results shown in the left panel of Figure \ref{fig:large_doubles_results} and Table 
\ref{tab:large_doubles_data}, we see that VMC is in closer agreement with coupled cluster, particularly 
CCSDT, instead of CASPT2.
This aligns with the conclusions of Loos and coworkers that the $ 2\ {}^1 A_g $ state has a high 
level of singles character, making coupled cluster with triples more reliable than it is for pure
double excitations.
In the case of the $ 2\ {}^1 A_{1g} $ state in benzene, the literature\cite{Loos2019} values 
consist only of different varieties of CASPT2 and NEVPT2, which show a significant spread of almost an
entire eV in their excitation energy predictions without an obvious reason to favor any particular value.
Our VMC result lands closest to the XMS-CASPT2 value, but is higher by about 0.1 eV and it may 
serve as a useful point of comparison for any other methods that are applied to this state in the 
future.

\subsection{Charge Transfer: Ammonia-difluorine}

Turning now to a benchmark charge transfer system, we test whether our
approach can maintain the high accuracy it has displayed in single
and double excitations.
The benchmark we use is the well-known ammonia-difluorine test system
\cite{Lucchese1975,Zhao2006,Dutta2018,Kozma2020}
at a separation of 6 \r{A}.
Our wave function generation proceeds as before with a (2e,6o) active space for the SS-CASSCF and a 
(22e,40o) space for the sCI using pseudopotentials and an aug-cc-pVTZ basis set.\cite{Bennett2017}
Note that augmentation in the basis is important here, as the charge
transfer essentially creates an anion.
For comparison, we performed $\delta$-CR-EOMCC(2,3)D\cite{Piecuch2015} calculations in 
GAMESS\cite{GAMESS2020} and EOM-CCSD and CASPT2 calculations in Molpro.\cite{MOLPRO_brief}

\begin{figure}[H]

\includegraphics[width=\columnwidth]{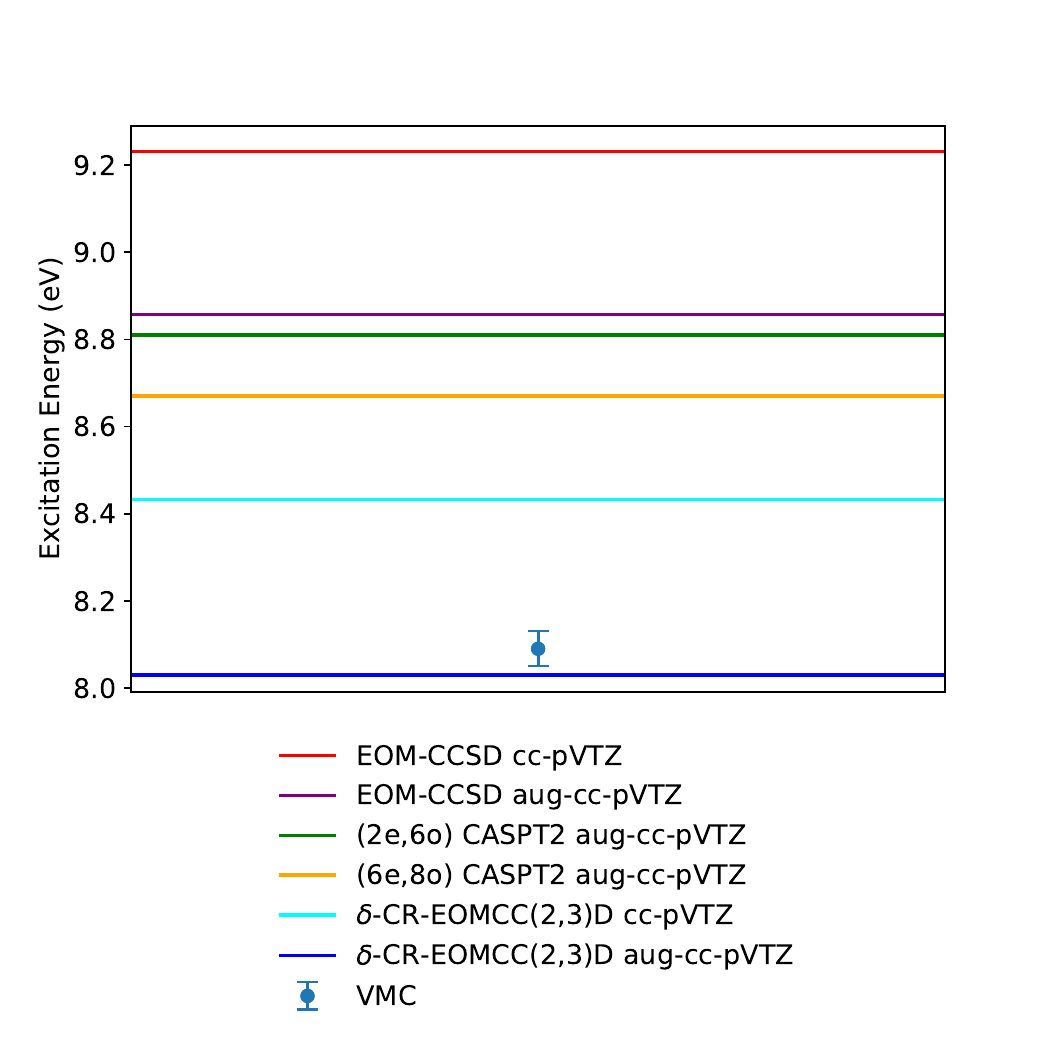}
\caption{Excitation energy of the lowest CT state in ammonia-difluorine for VMC and quantum chemistry approaches. See also Table \ref{tab:nh3-f2_data}. The VMC used 500 determinants for the excited state and interpolated between 5 and 300 for the ground state.}
\label{fig:nh3-f2_results}

\end{figure}

From Figure \ref{fig:nh3-f2_results} and Table \ref{tab:nh3-f2_data}, we see that our VMC is in 
excellent agreement with the aug-cc-pVTZ $\delta$-CR-EOMCC(2,3)D result.
With only a cc-pVTZ basis set, the $\delta$-CR-EOMCC(2,3)D excitation energy is 0.4 eV larger and 
EOM-CCSD has a substantial error of about 0.8 eV even with augmentation, showing
the importance of both including some triples in the coupled cluster and using diffuse functions for 
quantitative accuracy.
We also note that CASPT2 with a (2e,6o) active space,
a 6 state state-average (necessary to even see the CT state),
and IPEA shift of 0.25 has an error comparable to EOM-CCSD's, while 
increasing the active space to (6e,8o) with the same state averaging and shift improves the accuracy by only 0.14 eV. 
CASPT2's difficulties are likely due to the limitations of 
state-averaging for CT states, which we can avoid through our state-specific methodology.
The high accuracy of our VMC in this CT example indicates that we can obtain reliable excitation 
energies regardless of the type of excited state, a versatility that neither coupled cluster nor 
CASPT2 enjoys.

\begin{table}[H]
\footnotesize
\caption{Excitation energies for the CT state in ammonia-difluorine.The VMC uncertainty for the last digit is given in parentheses.}
\begin{tabular}{lS}
Method & \multicolumn{1}{l}{Excitation Energy (eV)}    \\ \hline

VMC    & 8.09(4)           \\
(2e,6o) CASPT2  aug-cc-pVTZ & 8.81                \\
(6e,8o) CASPT2  aug-cc-pVTZ & 8.67                \\
CCSD cc-pVTZ & 9.23            \\
CCSD  aug-cc-pVTZ & 8.86                \\
$\delta$-CR-EOMCC(2,3)D  cc-pVTZ& 8.43                \\
$\delta$-CR-EOMCC(2,3)D  aug-cc-pVTZ& 8.03                \\
\end{tabular}

\label{tab:nh3-f2_data}
\end{table}

\subsection{Larger Systems: Benzonitrile and Uracil}

\begin{figure}[H]

\includegraphics[width=\columnwidth]{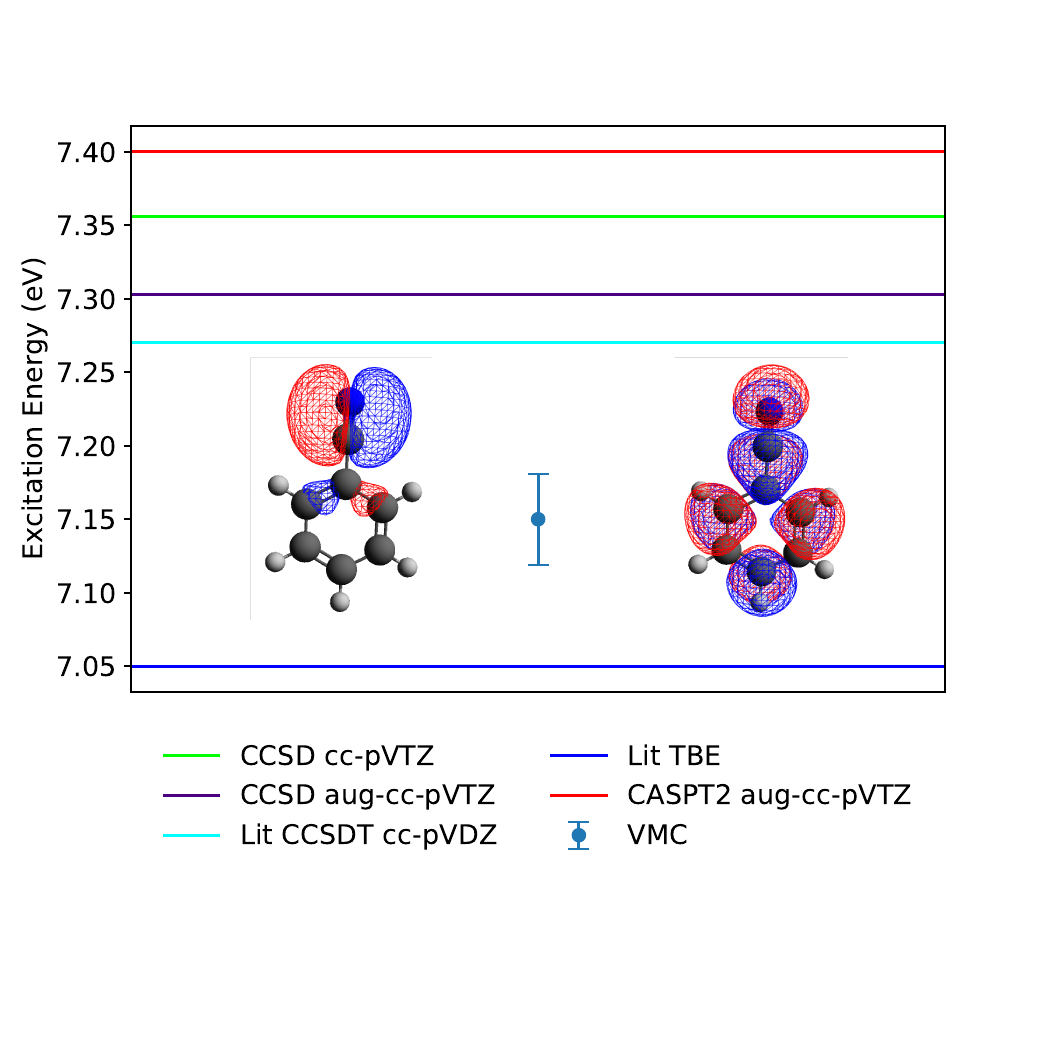}
\caption{Excitation energy of the CT state in benzonitrile for VMC and quantum chemistry approaches. See also Table \ref{tab:benzonitrile_data}. The VMC used 1000 determinants for the ground state and 200 for the excited state for an explicit variance match without interpolation. Benzonitrile is an atypical case where the ground state has the higher variance for fewer determinants. The SS-CASSCF hole and particle orbitals for the CT excitation are shown respectively on the left and right of the VMC result. Literature values taken from the work of Loos and coworkers.\cite{Loos2021}}
\label{fig:benzonitrile_results}

\end{figure}

We now turn our attention to somewhat larger molecules, where
brute-force reference values from exhaustive determinant
expansions in sCI are not currently possible.
We first address the CT state in benzonitrile, which is a $\pi_{CN}$ to $\pi^*$ excitation.\cite{Loos2021}
We used a (10e,6o) active space for the SS-CASSCF which we then
extended to (38e,50o) for the sCI, all using
pseudopotentials and an aug-cc-pVTZ basis set.\cite{Bennett2017}
We compare our results against a coupled cluster benchmark\cite{Loos2021} by Loos and coworkers and our 
own EOM-CCSD and CASPT2 calculations in Molpro.
We see in Figure \ref{fig:benzonitrile_results} that our VMC methodology is again very 
accurate in this larger CT system, within 0.1 eV of the
literature benchmark based on basis-set-corrected coupled cluster.
In contrast, state-averaged CASPT2
(4 states with a (10e,6o) space and 0.25 IPEA shift) again has
a more substantial error in modeling CT, although not as
severe as in the ammonia-difluorine system.

\begin{table}[H]
	
\footnotesize
\caption{Excitation energies for the CT state in benzonitrile. The VMC uncertainty for the final digits is given in parentheses. The literature TBE was obtained by applying basis set corrections based on CCSD, CC3, and CC2 to a CCSDT result in cc-pVDZ.\cite{Loos2021}}
\begin{tabular}{lS}

Method & \multicolumn{1}{l}{Excitation Energy (eV)}    \\ \hline

VMC    & 7.15(4)          \\
CASPT2  aug-cc-pVTZ & 7.40               \\
CCSD cc-pVDZ & 7.54            \\
CCSD cc-pVTZ & 7.36            \\
CCSD aug-cc-pVDZ & 7.40            \\
CCSD  aug-cc-pVTZ & 7.30                \\
Lit CCSDT cc-pVDZ & 7.27 \\
Lit TBE & 7.05

\end{tabular}

\label{tab:benzonitrile_data}
\end{table}

Our final excited states are the $ 2\ {}^1 A^{'} $ and $ 3\ {}^1 A^{'} $ states in uracil, which appear as part 
of the Thiel set\cite{Schreiber2008} and other benchmarking studies.\cite{Lorentzon1995,Kannar2014b}
To generate our ansatzes, we used a (10e,8o) active space for the SS-CASSCF and a (42e,42o) space for 
the sCI with pseudopotentials and a cc-pVTZ basis set.\cite{Bennett2017}
These states have been previously studied with different levels of coupled cluster and CASPT2, giving 
a significant range of answers that spans 0.7 eV for the $ 2\ {}^1 A^{'} $ state and almost an entire eV
for the $ 3\ {}^1 A^{'} $ state.
The latter state is notable for having a significant, though non-dominant, amount of double excitation 
character, in the range of 15 to 20 percent based on CASSCF analysis, which might lead to hesitation 
over whether to trust coupled cluster or CASPT2.
VMC is in a position to provide the necessary guidance, with our results ending up in good agreement 
with the CCSD(T) results of Szalay and coworkers\cite{Kannar2014b} for both states and more 
generally supporting the higher excitation energy predictions of coupled cluster over the CASPT2
results of Roos\cite{Lorentzon1995} and Thiel.\cite{Schreiber2008}
We also note that while the numbers of determinants needed in VMC after systematic expansion are 
larger than those used for smaller molecules, they remain modest compared
to the tens of thousands of determinants used on other recent VMC studies.

\begin{figure}[H]

\includegraphics[width=\columnwidth]{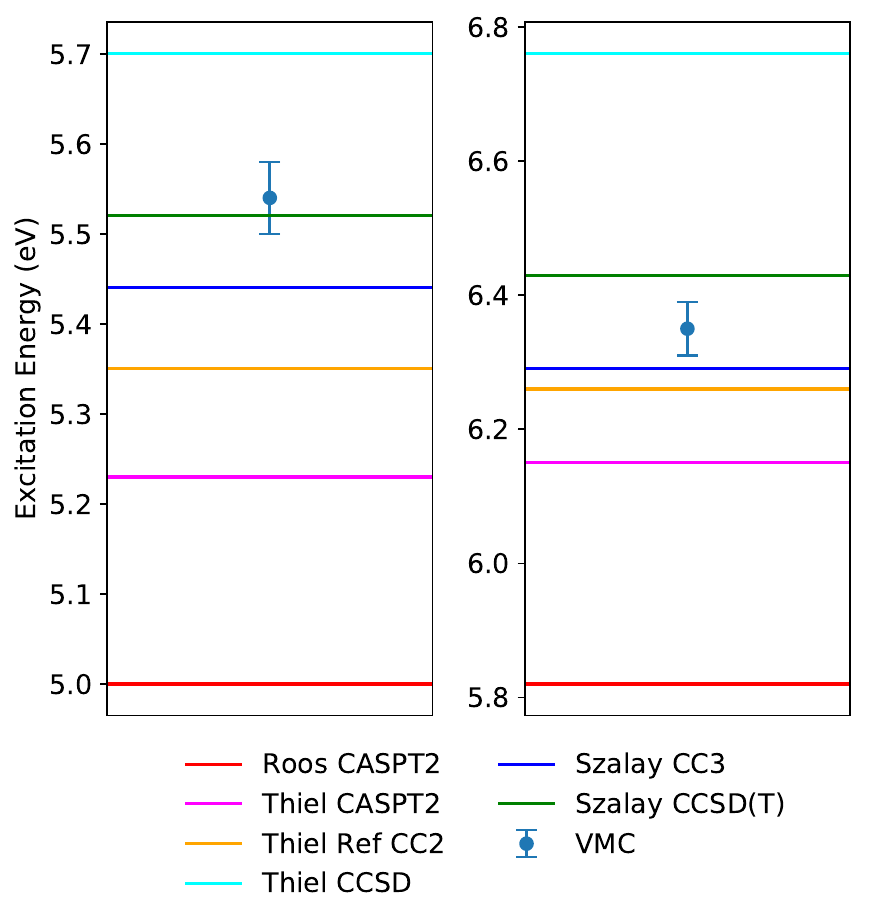}
\caption{Excitation energies in uracil. The left panel depicts values for the $ 2\ {}^1 A^{'} $ state and the right panel for the $ 3\ {}^1 A^{'} $ state. See also Table \ref{tab:uracil_data}. The VMC used 1500 determinants for each of the excited states and interpolated between calculations with 50 and 500 determinants for the ground state.}
\label{fig:uracil_results}

\end{figure}

\begin{table}[H]

\caption{Excitation energies in eV for the $ 2\ {}^1 A^{'} $  and $ 3\ {}^1 A^{'} $ states in uracil. The VMC uncertainty for the last digit is given in parentheses. The literature results from 
Thiel\cite{Schreiber2008} and Szalay\cite{Kannar2014b} used a TZVP basis set, with the Thiel set
using an aug-cc-pVQZ CC2 result as its reference. Details on the ANO type basis set used by Roos 
can be found in the original paper.\cite{Lorentzon1995}}
\begin{tabular}{lSS}
Method & \multicolumn{1}{c}{$ 2\ {}^1 A^{'} $}  & \multicolumn{1}{c}{$ 3\ {}^1 A^{'} $}  \\ \hline

VMC    & 5.54(4)        & 6.35(4)   \\

Roos CASPT2 & 5.00 &   5.82            \\
Thiel Ref CC2 & 5.35 &     6.26      \\
Thiel CCSD  & 5.70 &       6.76        \\
Thiel CASPT2  & 5.23 &      6.15         \\
Szalay CC3  & 5.44 &    6.29   \\
Szalay CCSD(T)  & 5.52 &    6.43           \\
\end{tabular}

\label{tab:uracil_data}
\end{table}

\section{Conclusions}

We have discussed how recent advances in excited state methods in the
fields of  quantum Monte Carlo and quantum chemistry have created
opportunities for fruitful method development at the intersection
of these two fields, in particular in the area of
excited-state-specific methods.
These opportunities are especially important in light of the
continued difficulty that established quantum chemistry methods
have in maintaining their accuracies across different classes
of excited states, with double excitations and charge transfer
excitations being two common spoilers.
Although this article has focused on one class of methods
that exist at this intersection, we prefer to see
these methods and the reliable accuracy they offer
primarily as motivations for further development at
this intersection rather than any sort of final word on what
is possible through combinations of excited-state-specific
quantum chemistry and quantum Monte Carlo.
Whole classes of methods that could contribute to this
fruitful intersection, such as excited-state-specific DFT
and diffusion Monte Carlo, bear further investigation,
and there are many more excited state situations to consider,
such as core states and Rydberg states, than we have discussed here.
We therefore see this intersection as offering a range of exciting
opportunities for further dramatic improvements in excited state simulation.

In the demonstrations we have included above, we showed how a
combination of SS-CASSCF, sCI, and excited-state-specific VMC
offers reliably accurate excitation energy predictions
across singly excited, doubly excited, and charge transfer states.
Unlike CASPT2, this approach avoids state averaging and is thus
able to fully accommodate the post-excitation orbital relaxations
that are important in charge transfer states.
Unlike EOM-CC, the approach is explicitly multi-configurational,
making it much more reliable in doubly excited states.
Unlike sCI and sCI-PT2 acting alone, the method is able to
arrive at accurate excitation energies with relatively short
determinant expansions thanks to the combined effect of Jastrow
factors and variance matching.
Together, these advantages lead to reliable excitation energy
predictions in a wider variety of excited states than either
coupled cluster or multi-reference perturbation theory while
reaching system sizes beyond the current limits of sCI.
In some systems, such as uracil and benzonitrile, agreement between this
approach and coupled cluster methods affords an especially high
degree of confidence in the accuracy of the results thanks to the
significant differences in these methods' underlying approximations.

From a methodological perspective, a key finding is that SS-CASSCF orbitals
are accurate enough that further optimization within VMC is not necessary
for producing accurate excitation energies.
The practical advantage of this finding is that, while more efficient than
it used to be, orbital optimization within VMC remains more difficult than
CI coefficient and Jastrow optimization, and by avoiding it we can therefore
reduce computational cost.
We also find that, in most cases, effective variance matching can be performed
without the need for interpolations based on multiple wave function expansion
lengths by instead using linear extrapolations based on optimization history.
Finally, removing low-priority parameters from the blocked linear method
steps while still optimizing them with accelerated descent did not prevent
high accuracies from being attained, which allows us to work with smaller
sample sizes than would otherwise be necessary for accurately estimating
the elements of the linear method matrices.


Looking forward, we are excited to explore wider applications of the
methodology presented here, as well as further method development at the
intersection between excited state quantum chemistry and quantum Monte
Carlo.
Core excitations are one area where VMC has already demonstrated high
accuracy in preliminary work, \cite{Garner2020a}
and we expect that the approach presented here will offer
similar advantages in that context.
Excited state absorption \cite{Fedotov2021} is another setting that challenges
both CASPT2 and coupled cluster methods but should pose much less
difficultly for methods that operate in an excited-state-specific mode.
Finally, the discrete states that exist at defects offer
opportunities for usefully applying state-specific methodology
in the solid state.
Methodologically, further improvements in user-accessibility are
an obvious priority, as much of the machinery behind modern excited
state VMC is relatively new and not yet familiar to many researchers.
Systematic determination and automation of effective choices in optimization
details such as the mixture between accelerated descent
and the blocked linear method steps in hybrid optimizations, as well as
the choice of guiding functions should help streamline workflows.
Most importantly, we fully expect that even more effective combinations
of methods at the intersection of excited state quantum chemistry
and quantum Monte Carlo will be developed in the years ahead.

\begin{acknowledgments}
This work was supported by the Office of Science, Office of Basic Energy Sciences, the U.S. 
Department of Energy, Contract No. DE-AC02-05CH11231.
Computational work was performed with the Berkeley Research Computing
Savio cluster, the LBNL Lawrencium cluster, and the National Energy Research Scientific Computing Center, a DOE Office of Science User Facility supported by the Office of Science of the U.S. Department of Energy under Contract No. DE-AC02- 05CH11231.
\end{acknowledgments}

\appendix

\section{Molecular Geometries}

  \begin{table}[H]
     \caption{Structure of thioformaldehyde. Coordinates in \r{A}.}
     \centering
     \begin{tabular}{lSSS}
       \multicolumn{1}{l}{} &  &  & \\ \hline
C   &   0.000000  & 0.000000 &  -1.104273 \\
S   &   0.000000  & 0.000000  &  0.514631 \\
H   &   0.000000  & 0.918957  & -1.677563 \\
H   &   0.000000  & -0.918957  & -1.677563 \\
       \hline
     \end{tabular}
     \label{tab:ch2sGeo}
   \end{table}

  \begin{table}[H]
     \caption{Structure of methanimine. Coordinates in \r{A}.}
     \centering
     \begin{tabular}{lSSS}
       \multicolumn{1}{l}{} &  &  & \\ \hline
C   &   0.056604  &  0.000000  &  0.587869 \\
N   &  0.056961   & 0.000000   & -0.686225 \\
H   & -0.842138   & 0.000000   & 1.202802 \\
H   &   1.007951  &  0.000000  &  1.108065 \\
H   &  -0.899369  &  0.000000  & -1.038338 \\
       \hline
     \end{tabular}
     \label{tab:ch3nGeo}
   \end{table}

  \begin{table}[H]
     \caption{Structure of ketene. Coordinates in \r{A}.}
     \centering
     \begin{tabular}{lSSS}
       \multicolumn{1}{l}{} &  &  & \\ \hline
C    &   0.000000  &  0.000000  & -1.295479 \\
C    &   0.000000   & 0.000000  &  0.018514 \\
O    &   0.000000   & 0.000000  &  1.183578 \\
H    &   0.000000   & 0.938930  & -1.818814 \\
H    &   0.000000   & -0.938930  & -1.818814 \\
       \hline
     \end{tabular}
     \label{tab:ch2coGeo}
   \end{table}

    \begin{table}[H]
     \caption{Structure of carbon trimer. Coordinates in \r{A}.}
     \centering
     \begin{tabular}{lSSS}
       \multicolumn{1}{l}{} &  &  & \\ \hline
C   &   0.000000  &  0.000000  &  0.000000 \\
C   &   0.000000  &  0.000000  &  1.298313 \\
C   &   0.000000  &  0.000000  & -1.298313 \\
       \hline
     \end{tabular}
     \label{tab:c3Geo}
   \end{table} 
  
    \begin{table}[H]
     \caption{Structure of nitrosomethane. Coordinates in \r{A}.}
     \centering
     \begin{tabular}{lSSS}
       \multicolumn{1}{l}{} &  &  & \\ \hline
C   &  -0.944193 &   0.000000  & -0.567405 \\
N   &  -0.002867 &   0.000000  &  0.571831 \\
O   &   1.157919  &  0.000000  &  0.229939 \\
H    & -0.409287 &   0.000000  & -1.515646 \\
H    & -1.574151 &   0.882677 &  -0.457339 \\
H    & -1.574151  & -0.882677  & -0.457339 \\
       \hline
     \end{tabular}
     \label{tab:ch3noGeo}
   \end{table}

   \begin{table}[H]
     \caption{Structure of hexatriene. Coordinates in \r{A}.}
     \centering
     \begin{tabular}{lSSS}
       \multicolumn{1}{l}{} &  &  & \\ \hline
C    & -0.000131  &  0.672998  &  0.000000 \\
C    &  0.000131  & -0.672998 &   0.000000 \\
C    &  1.201720  &  1.480569  &  0.000000 \\
C   &  -1.201720   & -1.480569  &  0.000000 \\
C    &  1.201720  &  2.821239  &  0.000000 \\
C   &  -1.201720  & -2.821239  &  0.000000 \\
H   &  -0.949240  &  1.200063  &  0.000000 \\
H   &   0.949240  & -1.200063  &  0.000000 \\
H   &   2.145737  &  0.948370  &  0.000000 \\
H   &  -2.145737  & -0.948370  &  0.000000 \\
H   &   0.274043  &  3.378403  &  0.000000 \\
H  &   -0.274043  & -3.378403  &  0.000000 \\
H  &    2.122561  &  3.385081  &  0.000000 \\
H   &  -2.122561  & -3.385081  &  0.000000 \\
       \hline
     \end{tabular}
     \label{tab:c6h8Geo}
   \end{table}
 
     \begin{table}[H]
     \caption{Structure of benzene. Coordinates in \r{A}.}
     \centering
     \begin{tabular}{lSSS}
       \multicolumn{1}{l}{} &  &  & \\ \hline
C   &   0.000000  &  1.392503  &  0.000000 \\
C   &  -1.205943  &  0.696252  &  0.000000 \\
C   &  -1.205943  & -0.696252  &  0.000000 \\
C   &   0.000000  & -1.392503  &  0.000000 \\
C   &   1.205943  & -0.696252  &  0.000000 \\
C   &   1.205943  &  0.696252  &  0.000000 \\
H   &  -2.141717  & 1.236521   &  0.000000 \\
H   &  -2.141717  & -1.236521  &  0.000000 \\
H   &   0.000000  & -2.473041  &  0.000000 \\
H   &   2.141717  & -1.236521  &  0.000000 \\
H   &   2.141717  &  1.236521  &  0.000000 \\
H   &   0.000000  &  2.473041  &  0.000000 \\
       \hline
     \end{tabular}
     \label{tab:c6h6Geo}
   \end{table}

  \begin{table}[H]
     \caption{Structure of ammonia-fluorine. Coordinates in \r{A}.}
     \centering
     \begin{tabular}{lSSS}
       \multicolumn{1}{l}{} &  &  & \\ \hline
N    &  0.000000  & -0.234913  & -3.520527 \\
H    &  0.000000  &  0.704739  & -3.904943 \\
H   &   0.813763  & -0.704739  & -3.904943 \\
H   &  -0.813763  & -0.704739  & -3.904943 \\
F   &   0.000000  & -0.234913  &  2.479473 \\
F   &   0.000000  & -0.234913  &  3.904943 \\
       \hline
     \end{tabular}
     \label{tab:nh3-f2Geo}
   \end{table}
   
      \begin{table}[H]
     \caption{Structure of benzonitrile. Coordinates in \r{A}.}
     \centering
     \begin{tabular}{lSSS}
       \multicolumn{1}{l}{} &  &  & \\ \hline
C   &   0.000000  &  0.000000  & -1.973927 \\
C   &   0.000000  &  0.000000  & -0.543573 \\
C   &   1.210548  &  0.000000  &  0.151304 \\
C   &  -1.210548  &  0.000000  &  0.151304 \\
C   &   0.000000  &  0.000000  &  2.234561 \\
C   &   1.206839  &  0.000000  &  1.539834 \\
C   &  -1.206839  &  0.000000  &  1.539834 \\
N   &   0.000000  &  0.000000  & -3.136116 \\
H   &   2.137508  &  0.000000  & -0.397299 \\
H   &  -2.137508  &  0.000000  & -0.397299 \\
H   &   2.140821  &  0.000000  &  2.076766 \\
H   &  -2.140821  &  0.000000  &  2.076766 \\
H   &   0.000000  &  0.000000  &  3.312081 \\
       \hline
     \end{tabular}
     \label{tab:c2f4-c2h4Geo}
   \end{table}
   
  \begin{table}[H]
     \caption{Structure of uracil. Coordinates in \r{A}.}
     \centering
     \begin{tabular}{lSSS}
       \multicolumn{1}{l}{} &  &  & \\ \hline
H   &  -2.025413  & -1.517742   & 0.000000 \\
H   &  -0.021861  &  1.995767  &  0.000000 \\
H   &   2.182391  & -1.602586  &  0.000000 \\
H   &  -0.026659  & -2.791719  &  0.000000 \\
C   &  -1.239290  &  0.359825 &   0.000000 \\
C   &   1.279718  & 0.392094  &  0.000000 \\
C   &   1.243729  & -1.064577  &  0.000000 \\
C   &   0.055755  & -1.709579  &  0.000000 \\
O   &  -2.308803  &  0.954763  &  0.000000 \\
O   &   2.287387  &  1.092936  &  0.000000 \\
N   &  -1.139515  & -1.026364  &  0.000000 \\
N   &   0.000000  &  0.978951  &   0.000000 \\
       \hline
     \end{tabular}
     \label{tab:uracilGeo}
   \end{table}

\clearpage
\onecolumngrid

 \clearpage  
\twocolumngrid

\nocite{*}
\bibliography{aipsamp}

\end{document}